\documentclass[12pt]{iopart}
\usepackage{color}

\usepackage{graphicx}
\begin{document}
\title{Coexistence of superconductivity and complex 4$f$  magnetism in Eu$_{0.5}$Ce$_{0.5}$BiS$_{2}$F}

\author{Hui-Fei Zhai, Pan Zhang, Zhang-Tu Tang, Jin-Ke Bao, Hao Jiang, Chun-Mu Feng, Zhu-An Xu and Guang-Han Cao}
\address{Department of Physics, Zhejiang University, Hangzhou 310027, China

Collaborative Innovation Centre of Advanced Microstructures, Nanjing 210093, China}

\ead{phyzhf@zju.edu.cn; ghcao@zju.edu.cn}

\begin{abstract}

EuBiS$_{2}$F is a self-doped superconductor due to the mixed valence of Eu. Here we report that, with the Ce substitution for Eu by 50 at.\%, the material exhibits ferromagnetic ordering at 8 K for the Ce-4$f$ moment, superconductivity at 2.2 K in the BiS$_2$ layers, and possibly antiferromagnetic ordering at 2.1 K for the Eu-4$f$ spins. The Eu valence is essentially divalent with the Ce incorporation. We tentatively interpret the coexistence of ferromagnetism and superconductivity by considering different Bi-6$p$ orbitals that are responsible for superconductivity itself and for mediating the ferromagnetic interaction, respectively. We argue that the antiferromagnetic ordering of the Eu-4$f$ spins is most likely due to a magnetic dipole-dipole interaction.

\ (Some figures may appear in colour only in the online journal)

\end{abstract}



\submitto{\JPCM}
\maketitle
\section{\label{sec:level1}Introduction}
The coexistence of superconductivity (SC) and magnetic long-range ordering (LRO) has been a long-standing issue in condensed matter physics\cite{maple,review,cao-minireview}. For a complex material containing both conduction electrons and local moments, SC may coexist with antiferromagnetic LRO, because superconducting Cooper pairs, with a typical size (i.e. superconducting coherence length) of hundreds of interatomic distance, 'feel' null net magnetic field. On the other hand. SC rarely coexists with \emph{ferromagnetic} LRO because the Cooper pairs' size is generally less than that of the magnetic domain, and SC is suppressed by the spontaneous magnetization and/or internal exchange field via an electromagnetic mechanism\cite{ginzburg} or a paramagnetic effect\cite{pme1,pme2}. Earlier examples of 'magnetic superconductors' mostly show coexistence of SC and antiferromagnetism (AFM), but SC is suppressed or quenched whenever ferromagnetism (FM) appears\cite{maple,review}. However, the EuFe${_2}$As${_2}$-related systems exceptionally show both SC for the Fe-3$d$ conduction electrons and FM for the Eu-4$f$ local moment\cite{Eu122P,Eu122cao,Eu122Co,jiao,Eu122Ir}. Recent neutron diffraction scattering\cite{jin1,nandi2,jin2} and resonant magnetic x-ray scattering\cite{nandi-xrs} confirmed Eu-4$f$ FM with an order moment of $\sim 7$ $\mu_{\mathrm{B}}/\mathrm{Eu}$. The coexistence of FM and SC (hereafter abbreviated as 'FM+SC') is qualitatively explained in terms of Fe-3$d$ multiband effect, which simultaneously enables SC in the FeAs layers and local-moment FM in the Eu sheets via an effective Ruderman-Kittel-Kasuya-Yosida (RKKY) interaction\cite{cao-minireview,Eu122cao}. The exchange coupling between Fe-3$d$ ($d_{yz}$ and $d_{zx}$) superconducting electrons and the Eu-4$f$ local moment can be very weak, as revealed by a recent time-resolved magneto-optical study\cite{optic}. Apart from the Eu-4$f$ FM, it was also reported that the Ce-4$f$ FM could coexist with SC in CeFe(As,P)O$_{0.95}$F$_{0.05}$ system\cite{Ce1111}.

Resembling to the doped EuFe${_2}$As${_2}$ systems, the Ce-containing BiS$_{2}$-layer based superconductor is possibly another example that hosts FM+SC\cite{Ce1121,Ln1121,Ce-awana,Ce-SX,CeOYM,Ce-neutron,Ce1121-XRA,paris,SrCeLYK}. As the first BiS$_{2}$-layer based material synthesized in 1970s\cite{CeBiSO}, CeBiS$_{2}$O intrinsically shows a semiconducting behaviour\cite{Ceparent,estructure}. Following the discovery of SC in LaBiS$_{2}$O$_{1-x}$F$_{x}$\cite{La1121}, SC at $T_{\mathrm{c}}\sim$3 K was realized via the fluorine doping in CeBiS$_{2}$O\cite{Ce1121}. The striking feature of the Ce-containing superconductor is the ferromagnetic-like transition at $T_{\mathrm{m}}\sim$ 5 K, which often prevails over the superconducting diamagnetism\cite{Ce1121,Ln1121,Ce-awana}. Later, a full phase diagram of CeBiS$_{2}$O$_{1-x}$F$_{x}$ was mapped out\cite{CeOYM}, which shows SC and FM in a broad region of 0.4 $< x\leq$ 1.0. Interestingly, both $T_{\mathrm{c}}$ and $T_{\mathrm{m}}$ achieve their maxima at 0.7 $\leq x\leq$ 0.9, implying a joint origin for them. The magnetism in the Ce sublattice is revealed to be an Ising FM by neutron scattering\cite{Ce-neutron}. X-ray absorption measurements\cite{Ce1121-XRA} indicate that the fluorine doping decreases the Ce valence, and the Ce valence is essentially 3+ in the coexistence regime. It is also revealed that the Ce-valence variation accompanies with a systematic changes in the atomic local structures\cite{paris}. Additionally, FM+SC has been found even in a Ce-diluted system Sr$_{0.5}$Ce$_{0.5}$FBiS$_2$\cite{SrCeLYK}. So far, the origin of FM+SC remains an open question.

EuBiS$_2$F is a newly discovered member in the BiS$_2$-based superconducting family, which exclusively shows a possible charge-density-wave (CDW) instability at 280 K and SC at 0.3 K without extrinsic doping\cite{Eu1121}. The uniqueness lies in the Eu mixed valence in connection with the crossing of Eu-4$f$ bands onto the Fermi level. At 1.8 K the Eu-4$f$ local moment possibly freezes into a spin-glass state. Since SC coexists with FM in Ce-diluted Sr$_{0.5}$Ce$_{0.5}$FBiS$_2$\cite{SrCeLYK}, it is of great interest to study the Eu$_{0.5}$Ce$_{0.5}$BiS$_{2}$F system to understand the interplay and compatibility between SC and the Ce/Eu-4$f$ magnetism. In this paper, we report the physical properties of Eu$_{0.5}$Ce$_{0.5}$BiS$_{2}$F polycrystalline samples. We found that the Ce-4$f$ moment orders ferromagnetically at $\sim$8 K, followed by a superconducting transition at 2.2 K. Meanwhile, a possible antiferromagnetic ordering of Eu-4$f$ spins appears at 2.1 K. Our results demonstrate that Eu$_{0.5}$Ce$_{0.5}$BiS$_{2}$F is a rare example in which SC simultaneously coexists with complex 4$f$ magnetism in a single material.

\section{Experimental details}
Polycrystalline samples of Eu$_{0.5}$Ce$_{0.5}$BiS$_{2}$F were synthesized by solid-state reactions under vacuum using powders of EuF$_{2}$ (Alfa Aesar, 99.9\%), Bi$_{2}$S$_{3}$ (99.995\%) and S (99.9995\%), and small pieces of Ce (99.8\%). The stoichiometric mixtures were loaded in an alumina
tube jacketed with an evacuated quartz ampoule which was then heated to 1000 K holding for 10 h. After cooling down, the mixtures were ground in an agate
mortar, and pressed into pellets under a pressure of 2000
kg/cm$^{2}$. The pellets were sintered at 1000 K for 10 h in vacuum
again. Note that all the procedures except for sample-heating and ampoule-sealing were conducted in a glove box filled with pure argon.

Powder x-ray diffraction (XRD) was performed at room temperature
using a PANalytical x-ray diffractometer (Model EMPYREAN) with a
monochromatic CuK$_{\alpha1}$ radiation. The lattice parameters were
obtained by least-squares fit with the correction of zero shift. Temperature-dependent resistivity was measured in a Cryogenic Mini-CFM measurement system by a standard four-terminal method. Gold wires were attached onto the samples' surface with silver paint. The size of the contact pads leads to a total uncertainty in the absolute values of resistivity of $\pm$15\%. Additional resistivity measurements down to 0.3 K were carried out in a $^3$He refrigerator inserted in a Oxford superconducting magnet system. Temperature-dependent of dc magnetization was performed on a Quantum Design Magnetic Property Measurement System (MPMS-5). The specific heat capacity
was measured down to 0.5 K using a relaxation technique, on a Quantum Design Physical Property Measurement System (PPMS-9).

\section{Results and discussion}

\begin{figure}
\center
\includegraphics[width=8cm]{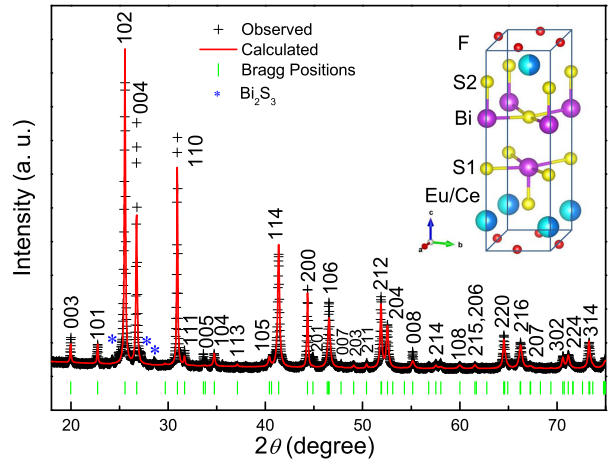}
\caption {\label{fig1:XRD} Powder x-ray diffraction of the Eu$_{0.5}$Ce$_{0.5}$BiS$_{2}$F sample. Three weak reflections marked by asterisks are identified to Bi$_2$S$_3$ impurity. Small amount of CeS impurity is also possible. The inset shows the crystal structure for the simulation of the XRD pattern.}
\end{figure}

\Fref{fig1:XRD} shows the powder XRD data of the as-prepared Eu$_{0.5}$Ce$_{0.5}$BiS$_{2}$F sample. The XRD pattern can be indexed by a tetragonal lattice with the space group \textit{P4/nmm} (No. 129). The unindexed reflections marked by asterisks are quite weak, which are identified to Bi$_2$S$_3$ impurity. We note that the Bragg-peak positions for a cubic CeS phase coincide with those of the main phase, and a multi-phase Rietveld refinement gives the amount of CeS of 3.3 at.\%. The unreacted impurity would lead to a slightly reduced Ce concentration of 0.48, compared with the nominal one. Note that both Bi$_2$S$_3$ and CeS impurities will not significantely influence the physical properties below because of their low fractions. The fitted lattice parameter, $a$ = 4.0697(1) {\AA}, is 0.47\% larger than that of its parent compound EuBiS$_{2}$F. However, another lattice parameter, $c$ = 13.3286(6) {\AA}, is 1.52\% smaller. As a result, the $c/a$ value has a significant decrease. The remarkable change in the lattice parameters indicates the incorporation of Ce, which induces electrons into the BiS$_2$ layers. Similar effect has been reported in many other BiS$_2$-based systems such as LaBiS$_{2}$O\cite{La1121}, CeBiS$_{2}$O\cite{Ce1121}, NdBiS$_{2}$O\cite{Nd1121}, and SrBiS$_{2}$F\cite{Sr1121}.

\begin{figure}
\center
\includegraphics[width=14cm]{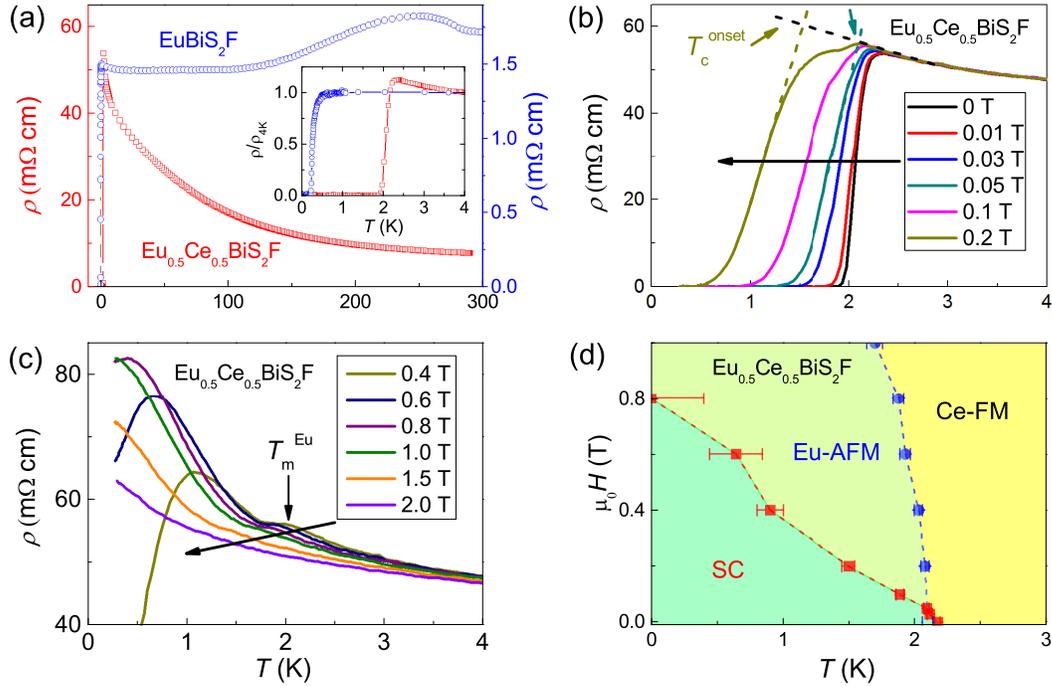}
\caption {\label{fig2:RT} (a) Temperature dependence of resistivity for Eu$_{0.5}$Ce$_{0.5}$BiS$_{2}$F (left axis) and EuBiS$_{2}$F (right axis) polycrystals. The inset highlights the superconducting transitions using a normalized resistivity. (b) and (c) The resistive superconducting transitions under various magnetic fields for Eu$_{0.5}$Ce$_{0.5}$BiS$_{2}$F. (d) The superconducting and magnetic phase diagram in Eu$_{0.5}$Ce$_{0.5}$BiS$_{2}$F.}
\end{figure}

Shown in \fref{fig2:RT}(a) is the temperature dependence of resistivity, $\rho(T)$, for the Eu$_{0.5}$Ce$_{0.5}$BiS$_{2}$F polycrystals. In contrast to EuBiS$_{2}$F, which exhibits metallic conduction and a CDW-like transition at 280 K, Eu$_{0.5}$Ce$_{0.5}$BiS$_{2}$F shows semiconducting-like behaviour with decreasing temperature before SC sets in. Note that $T_{\mathrm{c}}$ is enhanced by 7 times (from 0.3 K to 2.2 K), accompanied by disappearance of the CDW-like transition. The $T_{\mathrm{c}}$ enhancement can be understood by the additional electron doping with the Ce-for-Eu substitution. A very recent report shows the same $T_{\mathrm{c}}$ enhancement in La-doped EuBiS$_{2}$F\cite{EuLa1121}. Note that the semiconducting behaviour resembles those observed in other BiS$_2$-based materials\cite{Ce1121,La1121,Sr1121}, which is explained in terms of Anderson localization\cite{sakai}.

The resistive superconducting transitions under magnetic fields are shown in \fref{fig2:RT}(b) and (c). With increasing external magnetic fields, $T_{\mathrm{c}}$ decreases obviously and, the transitions become broader. Because of the broadened transitions as well as the semiconducting-like temperature dependence above $T_{\mathrm{c}}$, we define the onset transition temperature, $T_{\mathrm{c}}^{\mathrm{onset}}(H)$, by the crossing point of the extrapolation straight lines from the two sides (the normal and superconducting states). Consequently, the upper critical fields were obtained, as shown in the \fref{fig2:RT}(d). By using the Werthamer-Helfand-Hohenberg theory\cite{WHHtheroy} which applies to the orbital limited case, and the zero-temperature $\mu_{0}H_{\mathrm{c2}}^{\mathrm{orb}}(0)$ can be estimated to be $\sim$0.75 T, comparable to those in the FM-coexisted Sr$_{0.5}$Ce$_{0.5}$BiS$_{2}$F\cite{SrCeLYK} as well as nonmagnetic Sr$_{0.5}$La$_{0.5}$BiS$_{2}$F\cite{Sr1121}. The $\mu_{0}H_{\mathrm{c2}}^{\mathrm{orb}}(0)$ value is obviously lower than the Pauli paramagnetic limit $\mu_{0}$$H_{\mathrm{P}}$ = 1.84$T_{\mathrm{c}}$ $\approx$ 4 T. Since the Ce-4$f$ moment becomes ferromagnetically ordered at 8 K (see below), the unaffected upper critical field implies negligible coupling between the Cooper pairs and the Eu/Ce local moment.

There is a hump in $\rho(T)$ just above $T_{\mathrm{c}}(H)$ for $\mu_{0}H \geq 0.2$ T. This anomaly is gradually suppressed to lower temperatures, and then is smeared out with increasing magnetic fields. The hump is possibly due to an antiferromagnetic ordering of Eu spins (more evidences will be given by the magnetic and specific-heat measurements below). We also note that the resistivity decreases with increasing magnetic field below $\sim$ 10 K. The negative magnetoresistance in the normal state was also observed in other SC and FM coexisted BiS$_2$-based superconductors\cite{SrCeLYK}. Similar negative magnetoresistance was earlier reported in EuFe$_{2}$As$_{2-x}$P$_{x}$\cite{Eu122P} and CeFeAs$_{1-x}$P$_{x}$O$_{0.95}$F$_{0.05}$\cite{Ce1111}. The negative magnetoresistance comes from the ferromagnetic ordering for the Ce-4$f$ moment (see below).

\Fref{fig3:MT}(a) shows the dc magnetic susceptibility ($\chi$) measured under 1 kOe for Eu$_{0.5}$Ce$_{0.5}$BiS$_{2}$F. Considering the paramagnetic contributions from the local moments of both Eu and Ce ions, we fit the susceptibility by the formula, $\chi$ = $\chi_{0}$ + $C_{1}/(T - \theta_{1})$ + $C_{2}/(T - \theta_{2})$, where the $\chi_0$ is the temperature independent term, $C_{1}$ and $C_{2}$ are the Curie-Weiss constants of Eu and Ce ions, respectively, and $\theta_{1}$ and $\theta_{2}$ are their corresponding paramagnetic Curie-Weiss temperatures. The data fitting in a broad temperature range of 10 K $<T<$ 400 K without any parameter controlling gives the effective moments of $\mu_{\mathrm{eff}}$(Eu) = 8.7 $\mu_{\mathrm{B}}$/Eu and $\mu_{\mathrm{eff}}$(Ce) = 2.34 $\mu_{\mathrm{B}}$/Ce, which basically agrees with the expected values (7.94 $\mu_{\mathrm{B}}$/Eu$^{2+}$ and 2.54 $\mu_{\mathrm{B}}$/Ce$^{3+}$) for Eu$^{2+}$ and Ce$^{3+}$ ions. This result indicates that the Eu valence is essentially divalent, in contrast with the mixed valence in EuBiS$_{2}$F. In addition, the fitted $\theta_{1}$ and $\theta_{2}$ values ($-$3.7 and $+$6.5, respectively) are consistent with the effective \emph{antiferromagnetic} interactions between Eu$^{2+}$ ions and \emph{ferromagnetic} interactions between Ce$^{3+}$ ions.

\begin{figure}
\center
\includegraphics[width=8cm]{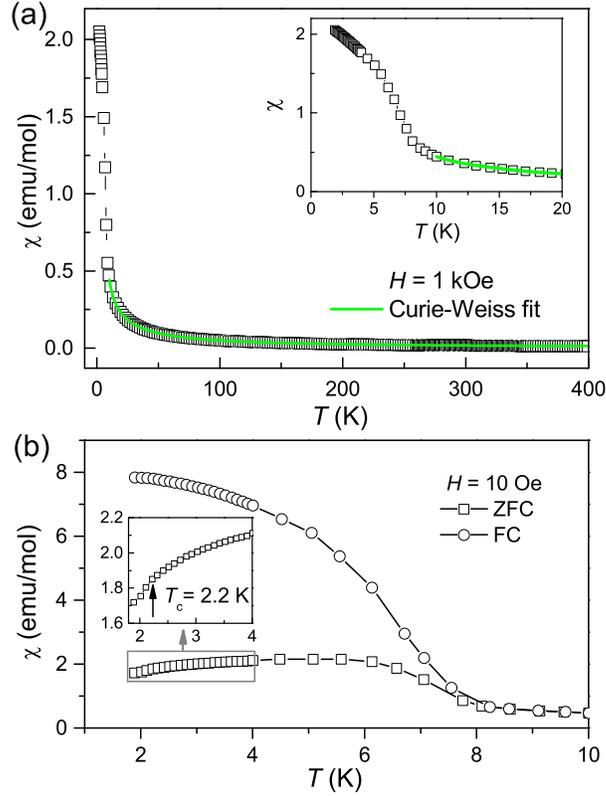}
\caption {\label{fig3:MT} Temperature dependence of dc magnetic susceptibility under a magnetic field of 1 kOe (a) and 10 Oe (b) for Eu$_{0.5}$Ce$_{0.5}$BiS$_{2}$F. The insets magnify the data at low temperatures. ZFC and FC denote zero-field cooling and field cooling, respectively.}
\end{figure}

\begin{figure}
\center
\includegraphics[width=8cm]{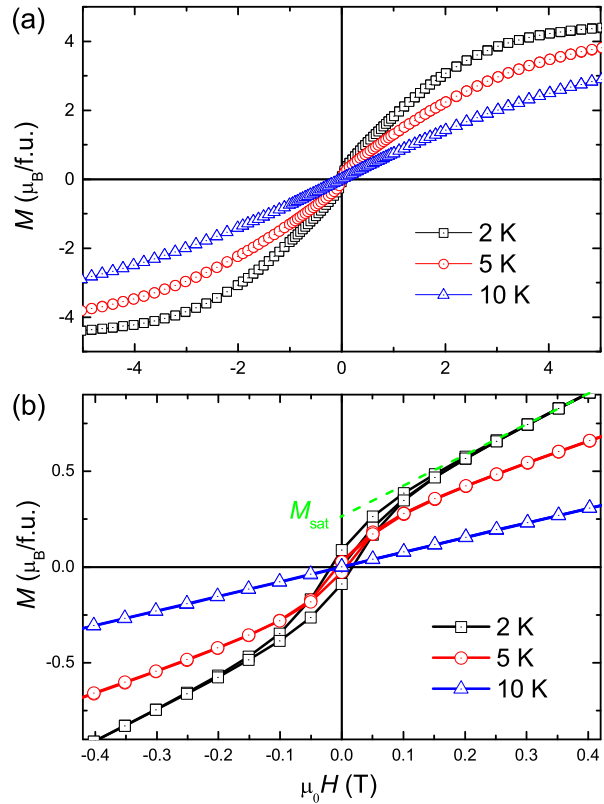}
\caption {\label{fig4:MH} (a) Field dependent magnetization at 2 K, 5 K and 10 K. (b) An enlarged plot showing the magnetic hysteresis due to the Ce-4$f$ ferromagnetic ordering.}
\end{figure}

Below 10 K, $\chi(T)$ increases rapidly with decreasing temperature, and it tends to saturate at lower temperatures, suggestive of a ferromagnetic transition. Another evidence of ferromagnetism is given by the divergence between zero-field-cooling (ZFC) and field-cooling (FC) susceptibility below 8 K, as shown in \fref{fig3:MT}(b). Indeed, the field dependence of magnetization, shown in \fref{fig4:MH}, displays a ferromagnetic-like hysteresis loop at 2 K, while such a hysteresis disappears at 10 K. The saturation magnetization $M_{\mathrm{sat}}$ is linearly extrapolated to be 0.27 $\mu_{\mathrm{B}}$/f.u. ('f.u.' denotes formula unit), corresponding to a 0.54 $\mu_{\mathrm{B}}$/Ce. This ordered magnetization is significantly lower than the expected value of 1.0 $\mu_{\mathrm{B}}$/Ce. There are several possible reasons. First of all, the $M_{\mathrm{sat}}$ could be underestimated due to (1) magnetocrystalline anisotropy, (2) the existence of SC, and (3) the Eu magnetism. The second reason comes from the possible deviation of Ce concentration. As mentioned above, the real Ce concentration could be $\sim$0.48 concerning the formation of CeS impurity. The last possibility is that the Ce moment actually forms ferromagnetic clusters under low external fields.

The magnetization continues to increase with increasing magnetic field, and it tends to saturate at 4.4 $\mu_{\mathrm{B}}$/f.u., owing to a field-induced Eu-spin reorientation. The final saturation magnetization is approximately equal to the sum of half an Eu$^{2+}$ spin (3.5 $\mu_{\mathrm{B}}$) and half a Ce$^{3+}$ moment (0.5 $\mu_{\mathrm{B}}$). Compared with the reduced magnetization of 5.58 $\mu_{B}$/Eu in EuBiS$_{2}$F due to the Eu mixed valence\cite{Eu1121}, the result further indicates that the Eu ions are divalent in Eu$_{0.5}$Ce$_{0.5}$BiS$_{2}$F.

The superconducting transition observed in the above resistivity measurement is not obvious for the magnetic data primarily because of the FM for Ce-4$f$ moment. There is no Meissner effect, and even the magnetic shielding effect is not certain, similar to the previous reports\cite{Ce1121,Ln1121,Ce-awana}. Here, as shown in \fref{fig3:MT}, a kink in the ZFC curve can be seen at 2.2 K, which could be due to the superconducting transition.

Figure 5 shows the temperature dependence of the specific heat, $C(T)$, for Eu$_{0.5}$Ce$_{0.5}$BiS$_{2}$F. The $C(T)$ curve tends to saturate to 130 J mol$^{-1}$ K$^{-1}$, basically consistent with the Dulong-Petit value 5 $\times$ 3$R$ $=$ 124.7 J mol$^{-1}$ K$^{-1}$, the high-\emph{T} limit for the lattice specific heat. The upper left inset magnifies the low-$T$ data. Two anomalies can be identified. The broad peak below 8 K is attributed to the ferromagnetic ordering of the Ce-4$f$ moment. The second peak centered at 1.5 K is probably associated with the magnetic ordering for the Eu$^{2+}$ spins. A similar peak was observed in EuBiS$_{2}$F, which was interpreted as a spin-glass transition because the peak is relatively broad and the magnetic entropy of the transition is reduced\cite{Eu1121}. However, the peak here is remarkably narrower, and the magnetic entropy essentially corresponds to the magnetic LRO scenario. The negative $\theta_{1}$ value as well as the feature of $M(H)$ curve at 2 K suggests antiferromagnetic order for the Eu$^{2+}$ spins. The superconducting specific-heat jump is hardly seen because of the reduced signal by the Ce-FM as well as the strong magnetic 'background' due to Eu-AFM.

\begin{figure}
\center
\includegraphics[width=8cm]{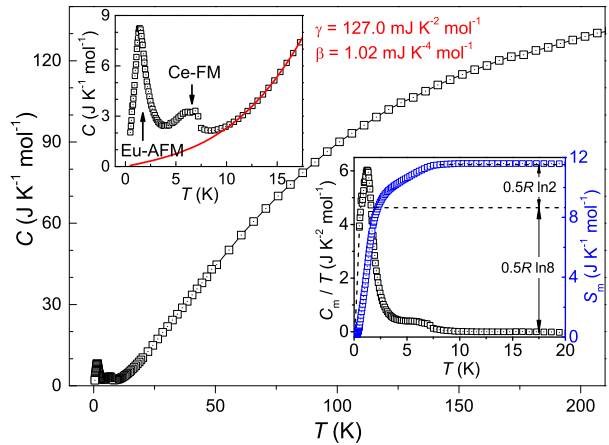}
\caption {\label{fig5:HC} Temperature dependence of specific heat for Eu$_{0.5}$Ce$_{0.5}$BiS$_{2}$F. The upper left inset plots the data below 20 K, in which the lattice and electronic contributions ($\gamma T$ + $\beta T^{3}$) are shown by the solid line. Ce-FM denotes ferromagnetism of Ce-4$f$ moment, and Eu-AFM represents antiferromagnetism of Eu-4$f$ spins. The lower inset shows the temperature dependence of $C_{\mathrm{m}}/T$ as well as the magnetic entropy $S_{\mathrm{m}}$.}
\end{figure}

The specific heat is generally contributed from several origins including crystalline lattice or phonons ($C_{\mathrm{lat}}$), conduction electrons ($C_{\mathrm{el}}$), and magnons ($C_{\mathrm{m}}$). The magnon contribution is expected to be negligible above 10 K, since no magnetic ordering takes place there. Then, we fit the $C(T)$ data in the temperature range of 10 K $<T<$ 18 K by employing the Debye $T^{3}$ law, $C = C_{\mathrm{el}} + C_{\mathrm{lat}} = \gamma T$ + $\beta T^{3}$. The Sommerfeld coefficient $\gamma$ is fitted to be 127.0 mJ mol$^{-1}$ K$^{-2}$, 1.7 times of that of EuBiS$_{2}$F (73.3 mJ mol$^{-1}$ K$^{-2}$)\cite{Eu1121}, 2.2 times of that of Sr$_{0.5}$Ce$_{0.5}$FBiS$_2$ (58.6 mJ mol$^{-1}$ K$^{-2}$)\cite{SrCeLYK}, and 90 times of that of Sr$_{0.5}$La$_{0.5}$FBiS$_2$ (1.42 mJ mol$^{-1}$ K$^{-2}$)\cite{Sr1121}. The great enhancement in Sommerfeld coefficient suggests the contributions owing to the Ce/Eu-4$f$ electrons. The fitting also yields $\beta$ = 1.02 mJ mol$^{-1}$ K$^{-4}$, from which the Debye temperature can be derived by the formula $\theta_{D}$ $=$ [(12/5)$NR$$\pi^{4}$/$\beta$]$^{1/3}$. The obtained Debye temperature (212.0 K) is reasonably higher than that of EuBiS$_{2}$F (201 K) and in between with those of LaBiS$_{2}$O$_{0.5}$F$_{0.5}$ and YbBiS$_{2}$O$_{0.5}$F$_{0.5}$\cite{Ln1121}.

The temperature dependence of the released magnetic entropy (from zero temperature) can be obtained by the integration $S_{\mathrm{m}}(T) = \int_{0}^{T}\frac{C_{\mathrm{m}}}{T}\mathrm{d}T$, where $C_{\mathrm{m}} = C - (\gamma T + \beta T^{3})$ (here we approximately assume $C_{\mathrm{m}}/T\propto T^{3/2}$ for 0 K $<T<$ 0.5 K). As shown in the lower inset of \fref{fig5:HC}, the released entropy is just equal to 50\% of $R\ln(2S + 1)$ ($S = 7/2$ for Eu$^{2+}$) at 2.3 K where the Eu-4$f$ spins become disordered. This strongly suggests that the Eu$^{2+}$ moment is fully ordered rather than partially ordered like that in a spin glass. When integrating to above the ferromagnetic transition temperature, additional 50\% of $R\ln(2J + 1)$ ($J = 1/2$ for Ce$^{3+}$) is released. This means that the Ce$^{3+}$ moment is also fully ordered at zero temperature. In short, our result indicates two types of magnetic LRO for the 4$f$ moments, FM for the Ce$^{3+}$ moment and possible AFM for the Eu$^{2+}$ spins, both coexisting with SC in Eu$_{0.5}$Ce$_{0.5}$BiS$_{2}$F.

\section{Discussion}

SC and FM are two antagonistic cooperative phenomena, and only in few special cases they can be compatible in a single material. So far, the uniform coexistence is believed to have realized exclusively in the uranium-based materials\cite{UGe2,URhGe,UCoGe}, in which both SC and FM come from the same type of electrons. The FM+SC in the EuFe${_2}$As${_2}$-related systems\cite{Eu122P,Eu122cao,Eu122Co,jiao,Eu122Ir} belongs to the case that SC and FM are from distinct electrons of different atoms, i. e. the Fe-3$d$ and Eu-4$f$ electrons. To be specific, SC is associated with the $d_{yz}$ and $d_{zx}$ orbitals of Fe 3$d$ electrons while FM arises from the LRO of Eu-4$f$ spins via the RKKY interactions mediated mainly through the $d_{x^2-y^2}$ and $d_{z^2}$\cite{Eu122cao}.

The Eu$_{0.5}$Ce$_{0.5}$BiS$_{2}$F bears similarities with the doped EuFe$_{2}$As$_{2}$ system, in which SC and FM emerge in different layers. Upon Ce doping, more electrons are doped into the BiS$_{2}$ layers, which leads to the enhancement of SC. The Bi 6$p$ orbitals mainly contribute to the electronic states near the Fermi level\cite{minimal model,wxg,yildirim}. Furthermore, the 6$p_{x}$ and 6$p_{y}$ electrons are believed to be responsible for SC\cite{minimal model}. It is expected that these superconducting electrons hardly couple with the 4$f$ electrons since the Bi 6$p_{x}$ and 6$p_{y}$ orbitals do not overlap with the 4$f$ orbitals located well above the BiS layers. This explains why SC is not influenced by the Ce/Eu-4$f$ local moments.

For the complex local-moment LRO, we note that the evolution of the magnetic ordering for the Ce-4$f$ moment and the Eu-4$f$ spin is very different. The Ce-FM is induced, accompanied with the emergence of SC, by the electron doping\cite{CeOYM,SrCeLYK}. Therefore, the Ce-FM is probably due to a RKKY exchange interaction in which the Bi-6$p_{z}$ electrons might play an important role. This scenario actually resembles the case in doped EuFe${_2}$As${_2}$, which simultaneously enables SC for 6$p_{x}$ and 6$p_{y}$ electrons in the BiS$_{2}$ layer and FM for the Ce-4$f$ moment mediated mainly by the 6$p_{z}$ itinerant electrons. For the Eu-AFM, on the other hand, the magnetic transition temperature, which is explicitly shown by the specific-heat peak, is essentially the same as that of the undoped EuBiS$_{2}$F\cite{Eu1121}. This suggests that the Eu-AFM is \emph{not} primarily due to a RKKY interaction. Considering the random occupation in the crystallographic site, such a magnetic LRO is not likely to happen through a superexchange interaction, either. Here we argue that the magnetic dipolar interaction could be the origin for the Eu-AFM. The nearest-neighbor dipolar coupling will be always antiferromagnetic for spins parallel to the $c$ axis, although Eu$^{2+}$ ions are randomly occupied in the lattice. Additionally, the high spins of Eu$^{2+}$ ions makes the magnetic interaction strong enough to form a magnetic order at about 2 K.

There are still some open questions. What is the magnetic interaction between Eu$^{2+}$ spins and Ce$^{3+}$ moment? What are the Eu/Ce spin directions in the lattice? Why Eu$^{2+}$ spins do not become ferromagnetic via a RKKY interaction, like the Ce$^{3+}$ moment?

\section{\label{sec:level5}Concluding remarks}
In summary, we demonstrate that Eu$_{0.5}$Ce$_{0.5}$BiS$_{2}$F is an interesting material which shows coexistence of superconductivity, ferromagnetism and possible antiferromagnetism. Superconductivity emerges in the BiS$_2$ layers, while the long-range magnetic ordering arises from the (Eu,Ce)F layers. Strikingly, the superconducting transition temperature as well as the upper critical field is hardly suppressed by the ferromagnetism for the Ce-4$f$ moment. By analogy with the scenario in EuFe$_{2}$As$_{2}$-related ferromagnetic superconductors\cite{cao-minireview,Eu122cao}, we tentatively interpret it by considering different Bi-6$p$ orbitals that are responsible for superconductivity itself and for mediating the RKKY interaction, respectively. The antiferromagnetism of the Eu-4$f$ spins is possibly due to a magnetic-dipole interaction. Further experimental and theoretical works are called for to confirm the interesting phenomenon and to clarify the underlying physics.

\ack  This work is supported by NBRPC (No. 2011CBA00103), NSFC (No. 11190023 and No. 11474252), and the Fundamental Research Funds for the Central Universities of China (No. 2013FZA3003).


\section*{References}

\begin{thebibliography}{00}
\bibitem{maple}Maple M B and Fisher {\O} 1982 {\it Superconductivity
in Ternary Compounds II} ({\it Topics in Current Physics} Vol 34) ed
M B Maple and {\O} Fisher (Berlin: Springer-Verlag) pp 1-10
\bibitem{review}Bulaevskii L N, Buzdin A I, Kulic M L and Panjukov S V 1985 {\it Adv. Phys.} \textbf{34} 175
\bibitem{cao-minireview}For a recent short review, see: Cao G H, Jiao W H, Luo Y K, Ren Z, Jiang S and Xu Z A 2012 {\it Journal of Physics: Conference Series} \textbf{391} 012123
\bibitem{ginzburg}Ginzburg V L 1956 {\it Zh. Eksp. Teor. Fiz.} \textbf{31} 202
\bibitem{pme1}Clogston A M 1962 {\it Phys. Rev. Lett.} \textbf{9} 266
\bibitem{pme2}Chandrasekhar B S 1962 {\it Appl. Phys. Lett.} \textbf{1} 7

\bibitem{Eu122P}Ren Z, Tao Q, Jiang S, Feng C M, Wang C, Dai J H, Cao G H and Xu Z A 2009 {\it Phys. Rev. Lett.} \textbf{102} 137002

\bibitem{Eu122cao} Cao G H, Xu S G, Ren Z, Jiang S, Feng C, and Xu Z A 2011 {\it J. Phys.: Condens. Matter} \textbf{23} 464204
\bibitem{Eu122Co}Jiang S, Xing H, Xuan G, Ren Z, Wang C, Xu Z A and Cao G H 2009 {\it Phys. Rev.} B \textbf{80} 184514
\bibitem{jiao}Jiao W H, Tao Q, Bao J K, Sun Y L, Feng C M, Xu Z A, Nowik I, Felner I and Cao G H 2011 {\it Europhys. Lett.} \textbf{95} 67007
\bibitem{Eu122Ir}Jiao W H, Zhai H F, Bao J K, Luo Y K, Tao Q, Feng C M, Xu Z A and Cao G H, 2013 {\it New J. Phys.} \textbf{15} 113002

\bibitem{jin1}Jin W T, Nandi S, Xiao Y, Su Y, Zaharko O, Guguchia Z, Bukowski Z, Price S, Jiao W H, Cao G H and Br\"{u}eckel Th 2013 {\it Phys. Rev.} B \textbf{88} 214516

\bibitem{nandi2}Nandi S, Jin W T, Xiao Y, Su Y, Price S, Schmidt W, Schmalzl K, Chatterji T, Jeevan H S, Gegenwart P and Br\"{u}eckel Th 2014 {\it Phys. Rev.} B \textbf{90} 094407
\bibitem{jin2}Jin W T, Li W, Su Y, Nandi S, Xiao Y, Jiao W H, Meven M, Sazonov A P, Feng E, Chen Y, Ting C S, Cao G H and Br\"{u}ckel Th 2015 {\it Phys. Rev.} B \textbf{91} 064506
\bibitem{nandi-xrs}Nandi S, Jin W T, Xiao Y, Su Y, Price S, Shukla D K, Strempfer J, Jeevan H S, Gegenwart P and Br\"{u}eckel Th 2014 {\it Phys. Rev.} B \textbf{89} 014512
\bibitem{optic}Pogrebna A, Mertelj T, Vuji\v{c}ic N, Cao G, Xu Z A and Mihailovic D 2015 {\it Sci. Rep.} \textbf{5} 7754
\bibitem{Ce1111}Luo Y K, Han H, Jiang S, Lin X, Li Y K, Dai J H, Cao G H and Xu Z A 2011 {\it Phys. Rev.} B \textbf{83} 054501

\bibitem{Ce1121}Xing J, Li S, Ding X, Yang H and Wen H H 2012 {\it Phys. Rev.} B \textbf{86} 214518
\bibitem{Ln1121}Yazici D, Huang K, White B D, Chang A H, Friedman A J and Maple M B 2012 {\it Phil. Mag.} \textbf{93} 673-680
\bibitem{Ce-awana}Jha R and Awana V P S 2014 {\it J. Supercond. Nov. Magn.} \textbf{27} 1-4
\bibitem{Ce-SX}Nagao M, Miura A, Demura S, Deguchi K, Watauchi S, Takei T, Takano Y, Kumada N and Tanaka I 2014 Solid State Communications \textbf{178} 33-36
\bibitem{CeOYM}Demura S, Deguchi K, Mizuguchi Y, Sato K, Honjyo R, Yamashita A, Yamaki T, Hara H, Watanabe T, Denholme S J, Fujioka M, Okazaki H, Ozaki T, Miura O, Yamaguchi T, Takeya H and Takano Y 2015 {\it J. Phys. Soc. Japan} \textbf{84} 024709
\bibitem{Ce-neutron}Lee J, Demura S, Stone M B, Iida K, Ehlers G, dela Cruz C R, Matsuda M, Deguchi K, Takano Y, Mizuguchi Y, Miura O, Louca D and  Lee S -H 2014 {\it Phys. Rev.} B \textbf{90} 224410
\bibitem{Ce1121-XRA} Sugimoto T, Joseph B, Paris E, Iadecola A, Mizokawa T, Demura S,  Mizuguchi Y,  Takano Y and Saini N L 2014 {\it Phys. Rev.} B \textbf{89} 201117(R)
\bibitem{paris}Paris E, Joseph B, Iadecola A, Sugimoto T, Olivi L, Demura S, Mizuguchi Y, Takano Y, Mizokawa T and Saini N L 2014 {\it J. Phys.: Condens. Matter} \textbf{26} 435701

\bibitem{SrCeLYK}Li L, Li Y K, Jin Y F, Huang H R, Chen B, Xu X F, Dai J H, Zhang L, Yang X J, Zhai H F, Cao G H and Xu Z A 2015 {\it Phys. Rev.} B \textbf{91} 014508

\bibitem{CeBiSO}Ceolin R and Rodier N 1976 {\it Acta Crystallogr. Sect.} B  \textbf{32} 1476
\bibitem{Ceparent}Higashinaka R, Asano T, Nakashima T, Fushiya K, Mizuguchi Y, Miura O, Matsuda T D and Aoki Y 2015 {\it J. Phys. Soc. Japan} \textbf{84} 023702
\bibitem{estructure}Morice C, Artacho E, Dutton S E, Molnar D, Kim H -J and Saxena S S 2013 arXiv:1312.2615
\bibitem{La1121}Mizuguchi Y, Demura S, Deguchi K, Takano Y, Fujihisa H, Gotoh Y, Izawa H and Miura O 2012 {\it J. Phys. Soc. Japan} \textbf{81} 114725


\bibitem{Eu1121}Zhai H F, Tang Z T, Jiang H, Xu K, Zhang K, Zhang P, Bao J K, Sun Y L, Jiao W H, Nowik I, Felner I, Li Y K, Xu X F, Tao Q, Feng C M, Xu Z A and Cao G H 2014 {\it Phys. Rev.} B  \textbf{90} 064518
\bibitem{Nd1121}Demura S, Mizuguchi Y, Deguchi K, Okazaki H, Hara H, Watanabe T, James Denholme S, Fujioka M, Ozaki T, Fujihisa H, Gotoh Y, Miura O, Yamaguchi T, Takeya H and Takano Y 2013 {\it J. Phys. Soc. Japan} \textbf{82} 033708

\bibitem{Sr1121}Lin X, Ni X X, Chen B, Xu X F, Yang X X, Dai J H, Li Y K, Yang X J, Luo Y K, Tao Q, Cao G H and Xu Z A 2013 {\it Phys. Rev.} B \textbf{87} 020504(R)



\bibitem{EuLa1121}Thakur G S, Jha R, Haque Z, Awana V P S, Gupta L C and Ganguli A K arXiv: 1504.08088
\bibitem{sakai}Sakai H, Kotajima D, Saito K, Wadati H, Wakisaka Y, Mizumaki M, Nitta K, Tokura Y and Ishiwata S 2014 {\it J. Phys. Soc. Japan} \textbf{83} 014709
\bibitem{WHHtheroy}Werthamer N R, Helfand E and Hohemberg P C 1966 {\it Phys. Rev.} \textbf{147} 295

\bibitem{UGe2}Saxena S S ,  Agarwal P,  Ahilan K, Grosche F M, Haselwimmer R K W,  Steiner M J, Pugh E, Walker I R, Julian S R, Monthoux P, Lonzarich G G, Huxley A, Sheikin I, Braithwaite D and Flouquet J 2000 {\it Nature} \textbf{406} 587
\bibitem{URhGe}Aoki D, Huxley A, Ressouche E, Braithwaite D, Flouquet J, Brison J P, Lhotel E and Paulsen C 2000 {\it Nature} \textbf{413} 613
\bibitem{UCoGe}Huy N T, Gasparini A, de Nijs D E, Huang Y, Klaasse J C P, Gortenmulder T, de Visser A, Hamann A, G\"{o}rlach T and L\"{o}hneysen H v 2007 {\it Phys. Rev. Lett.} \textbf{99} 067006

\bibitem{minimal model}Usui H, Suzuki K and Kuroki K 2012 {\it Phys. Rev.} B \textbf{86} 220501(R)
\bibitem{wxg}Wan X, Ding H -C, Savrasov S Y and Duan C G 2013 {\it Phys. Rev.} B \textbf{87} 115124
\bibitem{yildirim}Yildirim T 2013 {\it Phys. Rev.} B \textbf{87} 020506(R)
\end{thebibliography}
\end{document}